\documentstyle[twoside,fleqn,espcrc2,epsfig]{article}

\newcommand{\AmS}{{\protect\the\textfont2
  A\kern-.1667em\lower.5ex\hbox{M}\kern-.125emS}}

\def\ltap{\ \raisebox{-.4ex}{\rlap{$\sim$}} \raisebox{.4ex}{$<$}\ }
\def\gtap{\ \raisebox{-.4ex}{\rlap{$\sim$}} \raisebox{.4ex}{$>$}\ }
\title{New % Resonance 
      Enhancement Mechanism of % the Earth Effect in 
     the Transitions in the Earth
  of the Solar and Atmospheric Neutrinos Crossing the Earth Core}

\author{S.T. Petcov{\address{Scuola Internazionale Superiore di Studi Avanzati,
           and INFN (Trieste), I--34014 Trieste, Italy.}
       {\address{Inst. of Nuclear Research and Nuclear Energy, 
               Bulgarian Academy of Sciences, 1784 Sofia, Bulgaria.}}}}
       
\begin{document}
\begin{abstract}
It is shown that the $\nu_2 \rightarrow \nu_{e}$ and
$\nu_{\mu} \rightarrow \nu_{e}$ 
($\nu_e \rightarrow \nu_{\mu (\tau)}$)
transitions respectively of the solar and atmospheric neutrinos 
in the Earth in the case of  $\nu_e - \nu_{\mu (\tau)}$ % two-neutrino 
mixing in vacuum, are strongly enhanced by a new type of resonance
when the neutrinos cross the Earth core. 
The resonance is operative at small mixing angles 
but differs from the MSW one.
It is in many respects similar to the electron
paramagnetic resonance taking place 
in a specific configuration of two magnetic fields.
The conditions for existence of the new resonance
include, in particular, specific constraints on the neutrino oscillation
lengths in the Earth mantle and in the Earth core, thus the
resonance is a ``neutrino oscillation length resonance''.
It leads also to enhancement of the $\nu_2 \rightarrow \nu_e$ and
$\nu_e \rightarrow \nu_s$ transitions in the case of 
$\nu_e - \nu_s$ mixing and of the 
$\bar{\nu}_{\mu} \rightarrow \bar{\nu}_{s}$
(or $\nu_{\mu} \rightarrow \nu_{s}$) transitions
at small mixing angles. The presence of the neutrino oscillation length 
resonance in the transitions of solar and atmospheric neutrinos
traversing the Earth core has important implications for
current and future solar and atmospheric
neutrino experiments, and more specifically,
for the interpretation of the results of the
Super-Kamiokande experiment. 
\end{abstract}

\maketitle
\section{Introduction}
When the solar and atmospheric neutrinos traverse the Earth,
the $\nu_2 \rightarrow \nu_{e}$ and 
$\nu_{\mu} \rightarrow \nu_{e}$
($\nu_e \rightarrow \nu_{\mu(\tau)}$) transitions/oscillations
they undergo due to small $\nu_{\mu} - \nu_{e}$ mixing in vacuum
\footnote{As is well-known, the $\nu_2 \rightarrow \nu_{e}$
transition probability accounts for the Earth effect 
in the solar neutrino survival probability
in the case of the MSW two-neutrino  
$\nu_e \rightarrow \nu_{\mu(\tau)}$ and
$\nu_e \rightarrow \nu_{s}$ transition solutions of the
solar neutrino problem, $\nu_{s}$ being a sterile neutrino.}
can be dramatically 
enhanced by a new type of resonance
which differs from the MSW one and takes place 
when the neutrinos cross the 
Earth core \cite{SP3198}.
The resonance is present 
in the $\nu_2 \rightarrow \nu_{e}$ and 
$\nu_{\mu} \rightarrow \nu_{e}$
($\nu_e \rightarrow \nu_{\mu(\tau)}$) transition
probabilities, $P_{e2}$ and 
$P(\nu_{\mu (e)} \rightarrow \nu_{e (\mu;\tau)})$,
if the neutrino oscillation length (and mixing angles) in the 
Earth mantle and in the Earth core 
obey specific conditions \cite{SP3198}.
When satisfied, these conditions ensure that the 
relevant oscillating factors in the
probabilities $P_{e2}$ and 
$P(\nu_{\mu (e)} \rightarrow \nu_{e (\mu;\tau)})$
are maximal
\footnote{Note that, in contrast, the MSW effect is
a resonance amplifying the neutrino mixing.}
and that this 
produces a resonance maximum in $P_{e2}$ and 
$P(\nu_{\mu (e)} \rightarrow \nu_{e (\mu;\tau)})$.
Accordingly, the term ``neutrino oscillation 
length resonance'' % (NOLR) 
or simply ``oscillation length resonance'' was used 
in \cite{SP3198} to denote the resonance of interest. 
There exists a beautiful analogy between the neutrino 
oscillation length resonance and the electron 
spin-flip resonance realized  
in a specific configuration of 
magnetic fields \footnote{This analogy was brought 
to the attention of the author by L. Wolfenstein.} 
(see \cite{SP3198} for further details).

 At small mixing angles
($\sin^22\theta \ltap 0.05$)
the maxima due to the neutrino oscillation length resonance 
in $P_{e2}$ and $P(\nu_{\mu (e)} \rightarrow \nu_{e (\mu;\tau)}$) 
are absolute maxima and dominate in  $P_{e2}$ and 
$P(\nu_{\mu (e)} \rightarrow \nu_{e (\mu;\tau)})$:
the values of the probabilities at these maxima 
in the simplest case of two-neutrino mixing
are considerably larger -
by a factor of $\sim (2.5 - 4.0)$ ($\sim (3.0 - 7.0)$), 
than the values of $P_{e2}$ and 
$P(\nu_{\mu} \rightarrow \nu_{e}) = 
P(\nu_e \rightarrow \nu_{\mu(\tau)})$
at the local maxima associated with the MSW effect 
taking place in the Earth core (mantle).
The magnitude of the enhancement due to the 
oscillation length resonance % NOLR 
depends on the neutrino trajectory
through the Earth core: the enhancement is maximal for
the center-crossing neutrinos \cite{SP3198,s5398}. 
Even at small mixing angles 
the resonance is relatively wide both 
in the neutrino energy 
(or resonance density) \cite{SP3198} -
it is somewhat wider than the MSW resonance, 
and in the Nadir angle \cite{s5398}, $h$, 
specifying the neutrino 
trajectory in the Earth.
It also exhibits strong energy dependence.

  The presence of the oscillation length 
resonance in the transitions of   
solar and atmospheric 
neutrinos traversing the Earth 
has important implications \cite{SP3198,s5398,Art2,Art3,Art1} 
for the interpretation of 
the results, e.g., of the 
Super-Kamiokande experiment \cite{SKSuzukinu98,SKKajitanu98}.

 The Earth enhancement of the two-neutrino
transitions of interest
has been discussed rather extensively,
see, e.g., refs. \cite{Art1,PastAtmo}.
Some of the articles contain plots 
of the probabilities $P_{e2}$ and/or
$P(\nu_{\mu} \rightarrow \nu_{e})$ or
$P(\nu_e \rightarrow \nu_{\mu(\tau)})$   
on which one can 
recognize now the dominating maximum due to the neutrino 
oscillation length resonance (see, e.g., \cite{PastAtmo}). % ,PastSun}).
However, this maximum was invariably interpreted
to be due to the MSW effect in the Earth core
before the appearance of 
\cite{SP3198}.
\vspace*{-2pt}
\section{The Neutrino Oscillation Length Resonance (NOLR)}
\vspace*{-4pt}

  All the interesting features of the 
solar and atmospheric neutrino transitions 
in the Earth, including those related to 
the neutrino oscillation length resonance,
can be understood quantitatively 
in the framework of the two-layer model of the Earth density
distribution \cite{SP3198,s5398,3nuKP88,MP98:2layers}.
The density profile of the Earth in the two-layer model 
is assumed to consist of two structures - 
the mantle and the core, 
having different constant densities, $\bar{\rho}_{man}$ and $\bar{\rho}_{c}$, 
and different constant electron fraction numbers, 
$Y_e^{man}$ and $Y_e^{c}$
\footnote{The densities $\bar{\rho}_{man,c}$ 
should be considered as mean  
densities along the neutrino trajectories.
In the Earth model \cite{Stacey:1977} 
one has: $\bar{\rho}_{man} \cong (4 - 5)~ {\rm g/cm^3}$ and 
$\bar{\rho}_{c} \cong (11 - 12)~ {\rm g/cm^3}$.
For $Y_e$ one can use the standard 
values \cite{Stacey:1977,CORE} 
 (see also \cite{Art2}) $Y_e^{man} = 0.49$ and $Y_e^{c} = 0.467$.}.
The transitions of interest of the neutrinos 
traversing the Earth are essentially
caused by two-neutrino oscillations  
taking place i) first in the mantle over a distance $X'$
with a mixing angle $\theta'_{m}$ and oscillation length  
$L_{man}$, ii) then in the core over a 
distance $X''$ with different mixing angle $\theta''_{m}$ 
and oscillation length $L_{c}$, and iii) again
in the mantle over a distance $X'$
with $\theta'_{m}$ and $L_{man}$.  
Due to the matter effect $\theta'_{m},\theta''_{m}\neq \theta$
and $L_{man,c} \neq L_{vac}$, % $\theta$ and 
$L_{vac}$ being the
oscillation length in vacuum
(see, e.g., \cite{SPSchlad97}).
For fixed $X'$ and $X''$ 
the neutrino oscillation length 
resonance occurs \cite{SP3198}
if i) the relative phases acquired by the 
energy eigenstate neutrinos 
in the mantle and in the core,
$\Delta E'X'~ = 2\pi X'/L_{man}$ and $\Delta E''X'' = 2\pi X'' /L_{c}$,
are correlated, being odd multiples of $\pi$,  
so that % $L_{man,c}$ obey the constraints 
$$\frac{X'}{L_{man}} = k + \frac{1}{2},~~\frac{X''}{L_{c}} = k' + \frac{1}{2},
~~\eqno(1)$$
\noindent where $k,k' = 0,1,2,...$, and if ii) the inequality
$$\cos (2\theta''_{m} - 4\theta'_{m}  + \theta)~~(
\cos (2\theta''_{m} - 4\theta'_{m}))~ < 0 \eqno(2)$$
\noindent is fulfilled. Condition (2) is valid 
for the probability $P_{e2}$ ($P(\nu_{\mu} \rightarrow \nu_{e})$).
When equalities (1) hold, (2) ensures that $P_{e2}$ 
($P(\nu_{\mu} \rightarrow \nu_{e})$) has a maximum. 
In the region of the NOLR maximum where, e.g.,
$\Delta E'X' \cong \pi(2k + 1)$, 
$P_{e2}$ is given in the case of 
$\nu_{e} - \nu_{\mu}$ mixing by \cite{SP3198}:
$$P_{e2} \cong \sin^2\theta~+~ 
{1\over {4}} \left [1 - \cos \Delta E''X'' \right ]~~~~~~~~~~~~~~~~$$
$$~~~~~~~~~~\times~\left [ \sin^2(2\theta''_{m} - 4\theta'_{m}  + \theta) - 
 \sin^2\theta \right ].~\eqno(3)$$

\noindent At the NOLR maximum $P_{e2}$ takes the form \cite{SP3198}
$$P^{max}_{e2} = \sin^2(2\theta''_{m} - 4\theta'_{m}  + \theta).~\eqno(4)$$
\noindent The analogs of eqs. (3) - (4) 
for the probability $P(\nu_{\mu (e)} \rightarrow \nu_{e (\mu;\tau)})$ 
can be obtained by formally setting 
$\theta = 0$ while keeping $\theta_{m}' \neq 0$ and 
$\theta_{m}'' \neq 0$ in (3) - (4).
Note that one of the two NOLR requirements 
$\Delta E'X' = \Delta E''X'' = \pi$
is equivalent at $\sin^22\theta \ltap 0.02$ 
to the physical condition \cite{SP3198}
$$ \pi~({1\over {X'}} + {1\over {X''}})
\cong \sqrt{2} G_F (Y_e^{c}\bar{\rho}_{c} -
Y_e^{man}\bar{\rho}_{man}).~\eqno(5)$$

   Remarkably enough,  
for the $\nu_2 \rightarrow \nu_{e}$ and 
$\nu_{\mu (e)} \rightarrow \nu_{e (\mu;\tau)}$ transitions 
in the Earth, the NOLR conditions (1) with $k=k'=0$ 
are approximately fulfilled at small 
mixing angles ($\sin^22\theta \ltap 0.05$) 
in the regions where (2) holds \cite{SP3198}.
The associated NOLR maxima in  
$P_{e2}$ and $P(\nu_{\mu} \rightarrow \nu_{e})$ 
are absolute maxima (Figs. 1 - 2).

  Let us note that the study performed in \cite{SP3198}
and discussed briefly above 
\footnote{For % further details and 
analysis of the NOLR effects 
in the $\nu_{2} \rightarrow \nu_{e}$ 
and $\nu_e \rightarrow \nu_{s}$ 
transitions ($\nu_e - \nu_{s}$ mixing)
and in the $\bar{\nu}_{\mu} \rightarrow \bar{\nu}_{s}$
(or $\nu_{\mu} \rightarrow \nu_{s}$) 
transitions at small mixing angles see \cite{SP3198,s5398}.}
differs substantially from 
the studies \cite{Parametric}. The authors of
\cite{Parametric} considered the  
possibility of resonance enhancement of the 
$\nu_e \rightarrow \nu_{\mu (\tau)}$ transitions of 
neutrinos propagating in matter with density, varying 
{\it periodically} along the neutrino path 
(parametric resonance). It was found, in particular,
that at small mixing angles strong enhancement
is possible only if the neutrinos 
traverse at least 2 - 3 periods (in length)
of the density oscillations.
The density distribution in the
Earth is not periodic \footnote{The density change 
along the path of a neutrino crossing the Earth core is not
periodic even in the two-layer model: it falls short of making 
even one and a half periods.}; 
and in order for the oscillation length resonance 
\footnote{Although this term may not be perfect, 
it underlines the physical essence of the new resonance.  
The objection to it raised in \cite{ASnu98} 
is not convincing.
The term ``parametric resonance'' used in \cite{ASnu98}, e.g.,
suggests incorrect analogies.}
to occur periodic variation of the density is not required.
  
    In \cite{LAS98} the $\nu_{\mu} \rightarrow \nu_{s}$ 
transitions in the Earth were considered
for $\sin^22\theta \cong 1$. It was noticed that 
in the region where 
$\sqrt{2} G_F N_n^{man,c} % (1 - Y_e^{man,c})\bar{\rho}_{man,c} 
\gg \Delta m^2/E$, $N_n^{man,c}$ being the neutron number density, 
a new maximum in $P(\nu_{\mu} \rightarrow \nu_{s})$ appears when 
$\sqrt{2} G_F N_n^{man(c)} % (1 - Y_e^{man(c)})\bar{\rho}_{man(c)}
X'^{('')} \cong 2\pi$,
which was found to hold at $h \sim 28^{0}$. The height of the maximum 
is comparable to the heights of the other ``ordinary'' maxima 
present in $P(\nu_{\mu} \rightarrow \nu_{s})$ 
for $\sin^22\theta = 1$. It is % explicitly 
stated in \cite{LAS98} that the 
effect does not take place in the
$\nu_{\mu (e)} \rightarrow \nu_{e (\mu;\tau)}$ transitions, which is 
incorrect both for $\sin^22\theta \ll 1$ 
and $\sin^22\theta \cong 1$ \cite{SP3198,s5398}. 
\vspace*{-10pt}
\section{Implications of the Neutrino Oscillation Length Resonance}
\vspace*{-4pt}
  The implications of the oscillation length resonance 
enhancement of the probability $P_{e2}$ for the Earth core 
crossing solar neutrinos, for the tests of the MSW  
$\nu_e \rightarrow \nu_{\mu (\tau)}$ 
and $\nu_e \rightarrow \nu_{s}$  
solutions of the solar neutrino problem 
are discussed in refs. \cite{SP3198,s5398,Art2,Art3}. 
It is remarkable that for values of
$\Delta m^2 \cong (4.0 - 8.0) \times 10^{-6}~{\rm eV^2}$ 
from the small mixing angle (SMA) MSW solution region
and the geographical latitudes 
at which the Super-Kamiokande, SNO and ICARUS
detectors are located,
the enhancement takes place 
in the $\nu_e \rightarrow \nu_{\mu (\tau)}$ case for values of the 
$^{8}$B neutrino energy   
\begin{figure}[htb]
\vspace{8pt} % {9pt}
\framebox[73mm]{
 % {\rule[-21mm]{0mm}{43mm}}
\epsfig{file=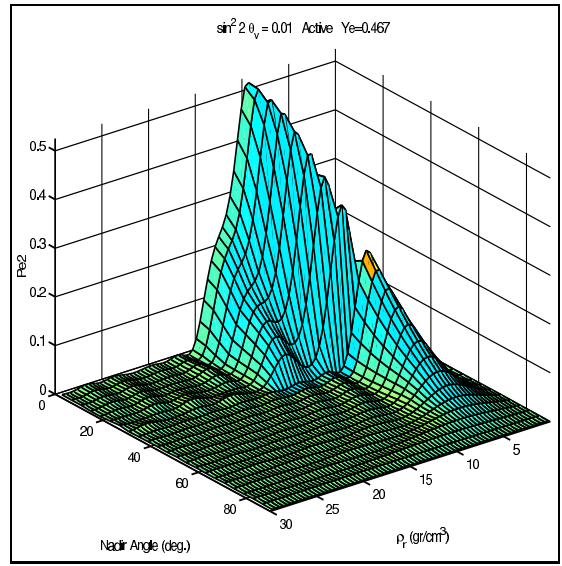,height=72mm,width=72mm}
}
\caption{
The probability $P_{e2}$ % ($\nu_e - \nu_{\mu (\tau)}$ mixing)
as a function of % the Nadir angle 
$h$ and the resonance density $\rho_r$ 
for $\sin^22\theta = 0.01$ \cite{s5398}. The prominent absolute maximum
for $h \approx (0^0 - 28^0)$ at $\rho_r \sim (8 - 10)~{\rm g/cm^3}$ 
is due to the NOLR. The ``shoulder'' at $\rho_r \sim 13~{\rm g/cm^3}$
is caused by the MSW effect in the Earth core \cite{SP3198}.}
\end{figure}
\noindent lying in the interval
$\sim (5 - 12)~$MeV to which 
these detectors are sensitive. The resonance maximum in
$P_{e2}$ at $\sin^22\theta = 0.01$
for the trajectory with % a Nadir angle 
$h = 23^{0}$, for instance,
is located at $E \cong 5.3~(10.5)~$MeV if 
$\Delta m^2 = 4.0~(8.0)\times 10^{-6}~{\rm eV^2}$.
Accordingly, at small mixing angles the NOLR
is predicted \cite{Art2} to produce  
a much bigger - by a factor of $\sim 6$, 
day-night (D-N) asymmetry in 
the Super-Kamiokande sample of solar 
neutrino events, 
whose night fraction is due to the core-crossing % solar 
neutrinos, in comparison with
the asymmetry determined by using the 
{\it whole night} event sample.
On the basis of these results it was concluded in \cite{Art2} that 
it can be possible to test a substantial part of the
MSW $\nu_e \rightarrow \nu_{\mu (\tau)}$ SMA 
solution region in the $\Delta m^2 - \sin^22\theta$
plane by   
\begin{figure}[htb]
% \vspace{0pt}  % {9pt}
\framebox[73mm]{
 % {\rule[-21mm]{0mm}{43mm}}
\epsfig{file=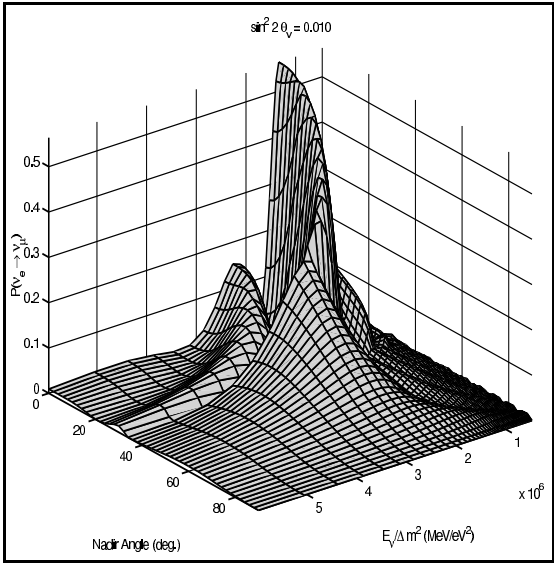,height=72mm,width=72mm}
}
\caption{The probability $P(\nu_{e (\mu)} \rightarrow \nu_{\mu;\tau (e)})$ 
as a function of $h$ and $E/\Delta m^2$ 
for $\sin^22\theta = 0.01$ \cite{s5398}. The absolute maximum
due to the NOLR for $ h \approx (0^0 -  28^0)$ is clearly seen 
at $E/\Delta m^2 \sim (1.3 - 1.6)\times 10^{6}~
{\rm MeV/eV^2}$. The local maximum at 
$E/\Delta m^2 \sim (2.5 - 3.0) \times 10^{6}~
{\rm MeV/eV^2}$ is due to the MSW 
effect in the Earth mantle.}
\end{figure}
\noindent performing {\it core} D-N asymmetry measurements.
The Super-Kamiokande collaboration
has already successfully
applied this approach to the analysis of their solar 
neutrino data \cite{SKSuzukinu98}:
the limit the collaboration has obtained on the
D-N asymmetry utilizing only the {\it core} 
event sample permitted to 
exclude a part of the MSW SMA solution region 
located in the area
$\sin^22\theta \cong (0.007 - 0.01)$,
$\Delta m^2 \cong (0.5 - 1.0)\times 10^{-5}~{\rm eV^2}$,
which is allowed by the mean event rate data 
from all solar neutrino experiments 
(Homestake, GALLEX, SAGE, Kamiokande and Super-Kamiokande). In contrast, 
the current Super-Kamiokande upper limit on 
the {\it whole night} D-N asymmetry \cite{SKSuzukinu98} 
does not permit to probe the SMA solution region: 
the predicted asymmetry is too small \cite{Art2}.

 The strong NOLR enhancement
of the $\nu_{\mu} \rightarrow \nu_{e}$ and 
$\nu_{e} \rightarrow \nu_{\mu (\tau)}$
transitions of atmospheric neutrinos 
crossing the Earth core
can take place at small mixing angles 
practically for all neutrino 
trajectories through the core \cite{SP3198}, e.g.,
for the trajectories with $h = (0^{0} - 23^{0})$ (Fig. 2). 
This is particularly relevant for the interpretation
of the results of the atmospheric neutrino experiments  
and for the future studies of the oscillations/transitions
of atmospheric neutrinos crossing the Earth.
The Super-Kamiokande collaboration 
has reported at this Conference strong 
evidences for oscillations
of the atmospheric $\nu_{\mu}$ ($\bar{\nu}_{\mu}$) 
\cite{SKKajitanu98}.
Assuming two-neutrino mixing, the data is best described   
in terms of $\nu_{\mu}(\bar{\nu}_{\mu}) 
\leftrightarrow \nu_{\tau}(\bar{\nu}_{\tau})$ 
% and $\bar{\nu}_{\mu} \leftrightarrow \bar{\nu}_{\tau}$
vacuum oscillations with parameters 
$\Delta m^2 \cong (0.5 - 6.0)\times 10^{-3})~{\rm eV^2}$
and $\sin^22\theta \cong (0.8 - 1.0)$.
The possibility of 
two-neutrino $\nu_{\mu}(\bar{\nu}_{\mu}) \leftrightarrow \nu_{e} (\bar{\nu}_{e})$  
large mixing oscillations 
is disfavored by the  
data \cite{SKKajitanu98}; at 
$\Delta m^2 \gtap 2\times 10^{-3}~{\rm eV^2}$ % this possibility 
it is ruled out \cite{CHOOZ98}. 

   It is a remarkable coincidence that for  
$\Delta m^2 \sim (0.5 - 6.0)\times 10^{-3})~{\rm eV^2}$
and small mixing, % angles, 
$\sin^22\theta \ltap 0.10$,
the oscillation length resonance 
in $P(\nu_{\mu} \rightarrow \nu_{e})
= P(\nu_{e} \rightarrow \nu_{\mu (\tau)})$
occurs \cite{SP3198} for values of the 
energy $E$ of the atmospheric 
$\nu_{e}$ and  $\nu_{\mu}$ 
which contribute either to the 
sub-GeV or to the multi-GeV 
$e-$like and $\mu-$like Super-Kamiokande event samples \cite{SKKajitanu98}.
For $\sin^22\theta = 0.01$,
$\Delta m^2 = 5\times 10^{-4};~10^{-3};~5\times 10^{-3}~{\rm eV^2}$, 
and $h = 0^{0}$ (Earth center crossing), 
for instance, the absolute maximum in 
$P(\nu_{\mu (e)} \rightarrow \nu_{e (\mu;\tau)})$ 
due to the NOLR 
takes place at $E \cong 0.75;~1.50;~7.5~$GeV.
Thus, for values of
$\Delta m^2$ % \sim (0.5 - 6.0)\times 10^{-3})~{\rm eV^2}$ of 
from the region of the $\nu_{\mu} \leftrightarrow \nu_{\tau}$
oscillation solution of the atmospheric neutrino problem, 
the NOLR  
strongly enhances the $\nu_{\mu} \rightarrow \nu_{e}$ (and 
$\nu_{e} \rightarrow \nu_{\mu (\tau)}$)
transitions of the atmospheric 
neutrinos crossing the Earth core, making 
the transitions 
detectable even at small mixing angles.
It was suggested in \cite{SP3198} % (see also \cite{SPNE98}) 
that the  
excess of e-like events
in the region $-1 \leq \cos\theta_{z}\leq -0.6$,
$\theta_{z}$ being the Zenith angle, either
in the sub-GeV or in the
multi-GeV sample, 
observed  (in both samples) in the
Super-Kamiokande experiment \cite{SKKajitanu98}, % SKAtmosph98}, 
is due to $\nu_{\mu} \rightarrow \nu_{e}$
small mixing angle transitions, $\sin^22\theta_{e\mu} \cong (0.01 - 0.10)$, 
with $\Delta m^2 \sim (0.5 - 1.0)\times 10^{-3})~{\rm eV^2}$ or
respectively $\Delta m^2 \sim (2 - 6)\times10^{-3})~{\rm eV^2}$, 
strongly enhanced by the NOLR 
\footnote{A more detailed investigation \cite{s5398,SPal98}
performed within the indicated three-neutrino mixing scheme
reveals, in particular, that the excess 
of e-like events in the Super-Kamiokande sub-GeV data 
at $-1 \leq \cos\theta_{z}\leq -0.6$
seems unlikely to be due to small mixing angle
% , $\sin^22\theta_{13} \ltap 0.20$, 
$\nu_{\mu} \rightarrow \nu_{e}$
transitions amplified by the oscillation length resonance.}.
The same resonantly enhanced transitions with 
$\Delta m^2 \sim (2 - 6)\times 10^{-3})~{\rm eV^2}$
($\Delta m^2 \sim (0.5 - 1.0)\times 10^{-3})~{\rm eV^2}$)
should produce 
at least part of the strong zenith angle dependence, 
exhibited by the $\mu-$like multi-GeV 
(sub-GeV) Super-Kamiokande data \cite{s5398}.

  The transitions of interest arise 
in a three-neutrino mixing scheme, in which 
the small mixing angle MSW 
$\nu_{e} \rightarrow \nu_{\mu}$ transitions 
with $\Delta m^2_{21} \sim (4 - 8)\times 10^{-6}~{\rm eV^2}$,
or large mixing angle 
$\nu_{e} \leftrightarrow \nu_{\mu}$ oscillations
with $\Delta m^2_{21} \sim 10^{-10}~{\rm eV^2}$,
provide the solution of the solar neutrino
problem and the atmospheric neutrino anomaly is
due to $\nu_{\mu} \leftrightarrow \nu_{\tau}$
oscillations with 
$\Delta m^2_{31} \sim (0.5 - 6.0)
\times 10^{-3}~{\rm eV^2}$ \cite{SP3198}.
For $\Delta m^2_{31} \gg \Delta m^2_{21}$ %, which seems to be 
the three-neutrino $\nu_{\mu} \rightarrow \nu_{e}$ 
and $\nu_{e} \rightarrow \nu_{\mu (\tau)}$
transition probabilities reduce \cite{3nuSP88} 
to the two-neutrino transition probability 
$P(\nu_{e} \rightarrow \nu_{\tau})$ (Fig. 2) 
with $\Delta m^2_{31}$  and 
$\sin^22\theta_{e\mu} =
4|U_{e3}|^2(1 - |U_{e3}|^2)$ playing 
the role of the 
two-neutrino oscillation parameters,
where $U_{e3}$ is the $e - \nu_3$ element of the lepton
mixing matrix, 
$\nu_3$ being the heaviest massive neutrino. 
The data \cite{SKKajitanu98,CHOOZ98} implies:
$\sin^22\theta_{13} \ltap 0.25$. Thus,
searching for the 
$\nu_{\mu} \rightarrow \nu_{e}$
and $\nu_{e} \rightarrow \nu_{\mu (\tau)}$ 
transitions of atmospheric neutrinos,
amplified by the oscillation length resonance,
can provide also unique information 
about the magnitude of  
$U_{e3}$ \cite{SPal98}.
\vspace*{-7pt}
\section{Conclusions}
\vspace*{-4pt}

  The neutrino oscillation length resonance should be
present in the $\nu_{2} \rightarrow \nu_{e}$
transitions taking place when the solar neutrinos
cross the Earth core on the way to the detector,
if the solar neutrino problem is due to 
small mixing angle MSW 
$\nu_{e} \rightarrow \nu_{\mu}$ 
transitions in the Sun.     
The same resonance should be operative also
in the $\nu_{\mu} \rightarrow \nu_{e}$
($\nu_{e} \rightarrow \nu_{\mu (\tau)}$) 
small mixing angle transitions 
of the atmospheric neutrinos crossing the Earth
core if the atmospheric $\nu_{\mu}$ and 
$\bar{\nu}_{\mu}$ indeed take part 
in large mixing vacuum 
$\nu_{\mu}(\bar{\nu}_{\mu}) \leftrightarrow \nu_{\tau}(\bar{\nu}_{\tau})$,
oscillations with 
$\Delta m^2 \sim (5\times 10^{-4} - 6\times 10^{-3})~{\rm eV^2}$,
as is strongly suggested by the Super-Kamiokande
data \cite{SKKajitanu98}, and if all 
three flavour neutrinos are mixed in vacuum.
The existence of three-flavour-neutrino mixing  
in vacuum is a very natural possibility in view of the 
present experimental evidences for oscillations/transitions of
the flavour neutrinos. In both cases the oscillation 
length resonance produces  
a strong enhancement of the 
corresponding transitions probabilities,
making the effects of the transitions observable 
even at rather small mixing angles. % \cite{SP3198}.
Actually, the resonance may have already
manifested itself in the excess of e-like events 
at $-1 \leq \cos\theta_{z}\leq -0.6$ observed
in the Super-Kamiokande multi-GeV atmospheric 
neutrino data \cite{SP3198,s5398,SPal98}.
And it can be responsible for at least part of the
strong zenith angle dependence present in the
Super-Kamiokande multi-GeV and sub-GeV $\mu-$like data \cite{s5398,SPal98}.

\end{document}